\documentclass[a4paper,11pt]{article}
\usepackage{jheppub} 
\usepackage{lineno}
\pdfoutput=1
\usepackage{amsmath,mathtools,amssymb,amsthm,amsxtra,overpic,bbm,epsfig,subfigure,url,bm}
\usepackage{hyperref}
\usepackage{mathrsfs}
\usepackage{color,xcolor}
\usepackage{comment}
\usepackage{float}
\usepackage{enumitem}
\usepackage{slashed}
\usepackage{fixmath}

\usepackage[T1]{fontenc}    
\usepackage{lmodern}        
\usepackage{graphicx}
\renewcommand\#{\protect\scalebox{0.8}{\protect\raisebox{0.3ex}{\char"0023}}}

\arxivnumber{2303.13706}

\title{Inferring astrophysical neutrino sources from the Glashow resonance}

\author{Guo-yuan Huang,}
\author{Manfred Lindner}
\author{and Nele Volmer}
\affiliation{Max-Planck-Institut f{\"u}r Kernphysik, Saupfercheckweg 1, 69117 Heidelberg, Germany}
\emailAdd{guoyuan.huang@mpi-hd.mpg.de, manfred.lindner@mpi-hd.mpg.de, nele.volmer@mpi-hd.mpg.de}
\abstract{We infer the ultrahigh energy neutrino source by using the Glashow resonance candidate event recently identified by the IceCube Observatory. For the calculation of the cross section for the Glashow resonance, we incorporate both the atomic Doppler broadening effect and initial state radiation $\overline{\nu}^{}_{e} e^- \to W^- \gamma$, which correct the original cross section considerably. Using available experimental information, we have set a generic constraint on the $\overline{\nu}^{}_{e}$ fraction of astrophysical neutrinos, which excludes the $\mu$-damped ${\rm p}\gamma$ source around $2\sigma$ confidence level under the assumption that neutrino production is dominated by the $\Delta$-resonance. While a weak preference has been found for the pp source, next-generation measurements will be able to distinguish between ideal pp and p$\gamma$ sources with a high significance assuming an optimistic single power-law neutrino spectrum.
The inclusion of multi-pion production at very high energies for the neutrino source can weaken the discrimination power. In this case additional multimessenger information is needed to distinguish between pp and p$\gamma$ sources.
}

\begin{document}

\maketitle
\flushbottom

\fontdimen1\font=0.0em
\fontdimen2\font=0.38em
\fontdimen3\font=0.1em
\fontdimen4\font=0.1em

\section{Introduction}The IceCube Observatory has successfully established the observation of ultrahigh energy (UHE) neutrino flux below a few PeV energies 
~\cite{IceCube:2021rpz,IceCube:2013low,IceCube:2013cdw,IceCube:2018cha,IceCube:2020abv,IceCube:2020wum,IceCube:2022der,Halzen:2022pez}. However, it remains a mystery as to where those neutrinos come from. 
One of the most popular mechanisms rests on accelerated cosmic rays colliding with  ambient targets around the source~\cite{Gaisser:1994yf,Bhattacharjee:1999mup,Beatty:2009zz,gaisser_engel_resconi_2016,anchordoqui2019ultra,tjus2020closing}.
There is a variety of source models for UHE neutrinos~\cite{Murase:2019tjj,Murase:2022feu,Troitsky:2021nvu,Xing:2011zza} which can usually be classified into the p$\gamma$ and pp types depending on whether the target particle is a photon or a proton.

For both p$\gamma$ and pp sources, after traveling an astronomical distance  the fluxes of three neutrino flavors strongly mix with each other due to neutrino oscillations, which ends up with a nearly democratic flavor composition $\phi^{\oplus}_{\nu_{e}} + \phi^{\oplus}_{\overline{\nu}_{e}} : \phi^{\oplus}_{\nu_{\mu}} + \phi^{\oplus}_{\overline{\nu}_{\mu}} : \phi^{\oplus}_{\nu_{\tau}} + \phi^{\oplus}_{\overline{\nu}_{\tau}} \approx 1:1:1$ at Earth~\footnote[4]{Throughout this work, we use the superscript `$\oplus$' to denote the quantity at Earth and `S' to denote that at source.}.
%
It is unlikely to disentangle those two sources by traditional flavor ratio measurements~\cite{Mena:2014sja, Chen:2014gxa, Palomares-Ruiz:2015mka, Aartsen:2015ivb, Palladino:2015zua, Arguelles:2015dca, Bustamante:2015waa, Aartsen:2015knd, Brdar:2016thq, DAmico:2017dwq, Pagliaroli:2015rca, Rasmussen:2017ert, Brdar:2018tce, Bustamante:2019sdb, Palladino:2019pid, Stachurska:2019srh,Song:2020nfh}.
The difference between those two sources lies in the composition of neutrinos and antineutrinos.
For the p$\gamma$ neutrino source, cosmic rays collide with photons to produce charged pions (mostly $\pi^+$ below a certain energy threshold) followed by the decays $\pi^+ \to \mu^+ \nu^{}_{\mu}$ and $\mu^+ \to e^+ \overline{\nu}^{}_{\mu} \nu^{}_{e}$, which results in more neutrino flux than antineutrino flux, i.e., $\phi^{\rm S}_\nu : \phi^{\rm S}_{\overline{\nu}} = 2:1$. In comparison, the pp source will give rise to nearly equal fractions of $\pi^+$ and $\pi^-$, which leads to $\phi^{\rm S}_\nu : \phi^{\rm S}_{\overline{\nu}} = 1:1$.

The key to distinguishing those two sources is by measuring the $\overline{\nu}^{}_{e}$ fraction $f^{}_{\overline{\nu}_{e}} \equiv \phi^{}_{\overline{\nu}_{e}}/(\phi^{}_{\overline{\nu}_{e}}+\phi^{}_{{\nu}_{e}})$, thanks to the Standard Model process $\overline{\nu}^{}_{e} e^- \to W^- \to X$ predicted by  S.~L.~Glashow~\cite{Glashow:1960zz}. 
Due to the resonance enhancement, the cross section of $\overline{\nu}^{}_{e} e^-$ scattering around $E^{}_{\nu} \approx 6.3~{\rm PeV}$ is larger than that of the deep inelastic scattering (DIS) by more than two orders of magnitude.
This promises us an excellent channel to differentiate between the ideal pp (with $f^{\oplus}_{\overline{\nu}_{e}} \approx 0.5$) and p$\gamma$ (with $f^{\oplus}_{\overline{\nu}_{e}} \approx 0.23$) sources, as continuously anticipated in previous works~\cite{Berezinsky:1977sf,Brown:1981ns,Anchordoqui:2004eb,Hummer:2010ai,Xing:2011zm,Bhattacharya:2011qu,Bhattacharya:2012fh,Barger:2012mz,Barger:2014iua,Palladino:2015uoa,Shoemaker:2015qul,Anchordoqui:2016ewn,Kistler:2016ask,Biehl:2016psj,Sahu:2016qet,Huang:2019hgs,Zhou:2020oym}.
It is worthwhile to emphasize that the above argument holds only for ideal  pp and p$\gamma$ cases. Other possible neutrino sources such as neutron and charm decays  can give rise to a $\overline{\nu}_{e}$ fraction different from the typical pion decays. Moreover, if the multi-pion channel in the p$\gamma$ scattering dominates the neutrino production when the collision energy is very high, more multimessenger information about the source in addition to the $\nu^{}_{e}$ fraction is necessary to disentangle the source degeneracy. 
In practice, the overall diffuse neutrino flux might be contributed by different types of sources, and the $\overline{\nu}_{e}$ fraction $f^{\oplus}_{\overline{\nu}_{e}}$ can take any reasonable values in between.
In this regard, we shall first treat the $\overline{\nu}_{e}$ fraction as a free parameter to be determined by the experimental probes regardless of the model assumptions. The experimental information about this parameter can then be used for theoretical interpretations under specific source assumptions.

Excitingly, with its unprecedented detection volume, the IceCube Observatory has collected one candidate event with an energy deposition $E^{}_{\rm dep} = 6.05 \pm 0.72~{\rm PeV}$ in the sample of partially contained events~\cite{IceCube:2021rpz}.
The probability that this event stems from the Glashow resonance (GR) is high, around $99\,\%$ by using the best-fit neutrino flux taken from ref.~\cite{IceCube:2015fuw}.
In this work, a timely quantitative assessment is carried out to infer the $\overline{\nu}^{}_{e}$ fraction by taking $f^{\oplus}_{\overline{\nu}_{e}} $ as a free parameter and to explain the level that we can differentiate between p$\gamma$ and pp sources. 
We have included both the radiation of initial photons~\cite{Gauld:2019pgt,Garcia:2020jwr,Alikhanov:2009vcg} and the Doppler broadening effect~\cite{Glashow:2014} while calculating the GR events.
Using the updated cross section, we investigate both the results for the current GR candidate in IceCube as well as the prospects of next-generation experiments.

\section{A full treatment of Glashow resonance}
As more and more UHE neutrino data have been accumulated, it becomes increasingly important to take into account the subleading effects for the theoretical evaluation of the GR. 
There are mainly two effects that should be emphasized: (i) the initial state radiation (ISR)~\cite{Gauld:2019pgt,Garcia:2020jwr}; (ii) the Doppler broadening effect~\cite{Glashow:2014}.
At the leading level, the 
cross section for the process $\overline{\nu}_e e^- \to W^- \to X$ reads \cite{IceCube:2021rpz}
\begin{equation}
\label{eq:sigmaGlashow}
\sigma^{(0)}(s)= 24 \pi {\rm \Gamma}^{2}_W {\rm Br}^{}_{W^-\to \overline{\nu}_e e^-}\frac{s/M_W^2}{(s-M_W^2)^2+{\rm \Gamma}_W^2M_W^2} \;,
\end{equation}
where $M_W \approx 80.433\, \text{GeV}$ is the mass of the $W$ boson, ${\rm \Gamma}_W \approx 2.09~{\rm GeV}$ is the total decay width and ${\rm Br}^{}_{W^-\to \overline{\nu}_e e^-} \approx 10.7\,\%$ is the branching ratio of the channel $W^-\to \overline{\nu}_e e^-$. 
The ISR and the Doppler broadening effect are found to considerably modify the above picture and should be included for completion.

Let us start with the ISR.
This effect becomes increasingly notable when the center-of-mass (COM) energy is much higher than the mass of the initial charged lepton, for which the collinear emission of photons is significant. 
For instance, in the Large Electron-Positron Collider (LEP), the ISR should be taken into account when analyzing the $Z$ boson peak~\cite{ALEPH:2005ab}.
For UHE neutrino telescopes like IceCube, the ISR cross section  near the GR will receive a large enhancement factor of  $\ln(M^{}_{W}/m^{}_{e}) \approx 12$ on top of the fine structure constant $\alpha$.

The ISR can be consistently included by using the structure function approach in analogy with the DIS off hadrons. The modified cross section will be \cite{Garcia:2020jwr}
\begin{equation}
\sigma (E^{}_{\nu})=\int \mathrm{d} x \, {\rm \Gamma}^{}_{e/e}(x,Q^2)\sigma^{(0)}(x,Q^2, E^{}_{\nu}) \;,
\end{equation}
where $Q$ represents the energy scale, $x$ is the longitudinal momentum fraction of the electron after the photon radiation, $\sigma^{(0)}$ is the cross section without the initial-state photon, and ${\rm \Gamma}^{}_{e/e}$ is the structure function of the electron.
We take the structure function from ref.~\cite{Cacciari:1992} which includes soft photons resummed to all orders and hard photons up to $\mathcal{O}(\alpha^3)$.

\begin{figure} 
	\centering
	\includegraphics[width=0.7\textwidth]{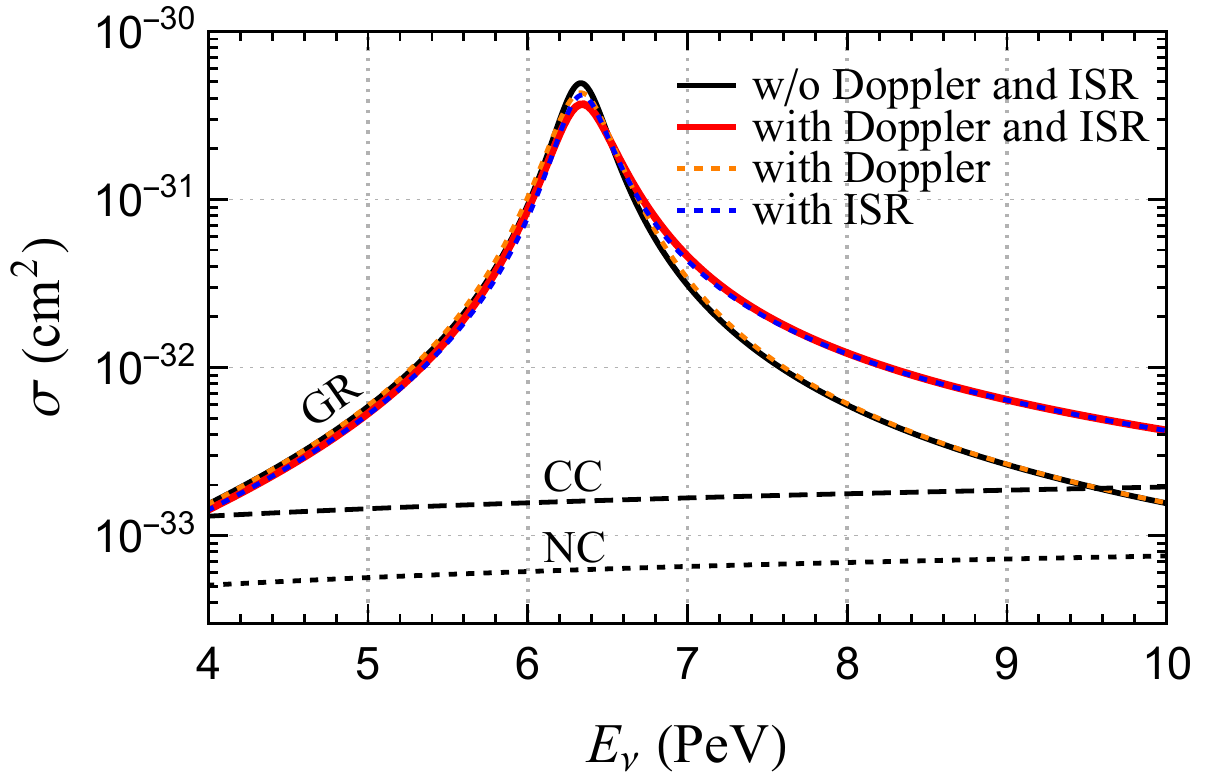}
	\caption{Cross section for the Glashow resonance process $\overline{\nu}_e+e^- \rightarrow W^- \rightarrow X$ with and without the initial state radiation and Doppler broadening effect. The black curve shows the cross section without initial state radiation and Doppler broadening, the blue dotted one includes initial state radiation and the orange dotted one includes Doppler broadening. The red curve is the cross section with both Doppler broadening and initial state radiation effects, and the tabulated result of this curve is given in our supplementary material. Both the broadening and the radiative return are visible. For the Glashow resonance curves we averaged over the electrons in H$_2$O for the target. }
	\label{fig:ComparisonAll}
\end{figure}

The second effect of interest is the Doppler broadening due to the motion of atomic electrons~\cite{Glashow:2014}. 
The velocity of atomic electrons $\beta$ is typically of the order $\mathcal{O}(\alpha\, c)$.
A simple estimation shows that this velocity will shift the COM energy square from $s = 2 E^{}_{\nu} m^{}_{e}$ to $2 E^{}_{\nu} m^{}_{e}(1-\beta\cos \theta)$, where $\theta$ is the angle between the electron velocity and the incoming neutrino in the laboratory frame. This broadens the COM energy by around $0.6~{\rm GeV}$ in comparison to the $W$ decay width ${\rm \Gamma}_W=~2.09~{\rm GeV}$.
Non-relativistic electrons in the atom have the four-momentum
$
( m_e+{\left|\mathbold{k}\right|^2}/({2m_e}),\mathbold{k})
$,
where $\left|\mathbold{k}\right| \approx m_e \beta$. By integrating over the electron wave function, one can arrive at the total cross section~\cite{Glashow:2014}
\begin{equation} 
\label{eq:integrationDoppler}
\sigma (E^{}_{\nu})= \frac{1}{4 \pi}\int \mathrm{d}\phi \int \mathrm{d}\beta F(\beta) \int \mathrm{d} x^\prime \sigma^{(0)}[E^{}_{\nu}(1-\beta x^\prime)] \;,
\end{equation}
where $\phi$ represents the azimuth angle, $F(\beta)$ is the velocity distribution of electrons and  $x^\prime=\cos (\theta )$.
Since the calculation framework was already outlined in ref.~\cite{Glashow:2014}, we give more details about the updated calculation in appendix~\ref{AppendixDoppler}.

Those two effects can be combined, and their joint result  is shown as the red curve in fig.~\ref{fig:ComparisonAll} for the ${\rm H_2 O}$  target, along with the cross sections without (solid black curve) or modified by only one  (blue and orange curves) of those effects. In comparison, the charged-current (CC) and neutral-current (NC) interactions are depicted as dashed and dotted black curves, respectively.
Some remarks on the results are given below.
\begin{itemize}[noitemsep,topsep=0pt,leftmargin=5.5mm]
	\item The ISR will reduce the peak at the resonance energy $E^{}_{\nu} \approx 6.3~{\rm PeV}$ by almost $20\,\%$. Furthermore, the cross section above the resonance energy is enhanced by a factor of more than two. This is due to the radiative return phenomenon, for which the photon in the process $\overline{\nu}^{}_{e} e^- \to W^- \gamma$ carries away some energy such that the $W$ production will be made on shell even if~$\sqrt{s} > M^{}_{W}$.
	\item The Doppler broadening effect for the ${\rm H_2 O}$  target is small compared to the ISR in the logarithmic scale.
	To see the detailed impact we also show the result in a flat scale as fig.~\ref{fig:DopplerBroadened}. 
	The resonance peak is reduced slightly, while the width is broadened due to the motion of atomic electrons.
	\item The combined result of the ISR and the Doppler broadening is obtained with a convolution, which reduces the peak by around $30\,\%$. However, we should note that those effects will be partly smeared by the finite energy resolution of the IceCube detector. We have checked that the eventual effect can decrease the events within the energy window near the GR by almost $10\%$.
\end{itemize}
With the full GR cross section, we are able to calculate the event rate in IceCube and compare it to both experimental data available now and those from future experiments.

\section{Analysis framework}%
In order to constrain the $\overline{\nu}^{}_{e}$ fraction in the total diffuse neutrino flux, we calculate the likelihood by fitting models with different values of $f^{}_{\overline{\nu}_{e}} = \phi^{}_{\overline{\nu}_{e}}/(\phi^{}_{\overline{\nu}_{e}}+\phi^{}_{{\nu}_{e}})$ to the available IceCube data.
The reason why we use $f^{}_{\overline{\nu}_{e}}$ to measure the $\overline{\nu}^{}_{e}$ fraction is that it almost solely determines the spectrum of single cascade event topology at PeV energies in the IceCube detector.

The observed GR candidate in IceCube belongs to the PeV energy partially contained events (PEPEs), in comparison to the high energy starting events (HESEs) where the shower is fully contained inside the fiducial volume.
Even though the PEPE effective volume is nearly twice the volume of HESE at PeV energies
only one event with an energy deposition $E^{}_{\rm dep} = 6.05 \pm 0.72~{\rm PeV}$ has been observed within the energy window $4~{\rm PeV} < E^{}_{\rm dep} < 10~{\rm PeV}$.
For HESE three PeV events have been collected~\cite{IceCube:2014stg}, nicknamed Bert, Ernie and Big Bird. However, all of them have energies below $3~{\rm PeV}$, which are most likely contributed by the DIS.
Even though the GR has not significantly arisen in the HESE sample, HESE is useful to fix the normalization and shape of UHE neutrino flux which are crucial for our extraction of the $\overline{\nu}^{}_{e}$ fraction.

In ref.~\cite{IceCube:2020wum}, the IceCube Collaboration has analyzed the overall UHE neutrino flux with HESEs collected over 7.5 years, assuming a flavor ratio $\phi^{\oplus}_{\nu_{e}} + \phi^{\oplus}_{\overline{\nu}_{e}} : \phi^{\oplus}_{\nu_{\mu}} + \phi^{\oplus}_{\overline{\nu}_{\mu}} : \phi^{\oplus}_{\nu_{\tau}} + \phi^{\oplus}_{\overline{\nu}_{\tau}} = 1:1:1$. 
During our analysis we will use the HESE results including uncertainties from ref.~\cite{IceCube:2020wum} to set the spectrum of neutrino flux and use  PEPE to extract the $\overline{\nu}^{}_{e}$ fraction $f^{\oplus}_{\overline{\nu}_{e}}$.
Note that a more thorough analysis would assume a completely free flavor ratio. However, on the one hand, the latest IceCube HESE fit available has fixed the flavor ratio \cite{IceCube:2020wum}. On the other hand, ideal pp and p$\gamma$ astrophysical models reasonably prefer such a democratic ratio after neutrino oscillations over an astronomical distance.

For demonstration, we choose two benchmark flux models in our analysis: (i) the unbroken single power-law model; (ii) the single power-law model with an exponential energy cutoff. The former one reads
\begin{align} \label{eq:dPhidE}
\frac{\mathrm{d} {\rm \Phi}^{}_{6\nu}}{\mathrm{d} E^{}_{\nu}} = {\rm \Phi}^{}_{0} \left(\frac{E^{}_{\nu}}{100~{\rm TeV}}\right)^{-\gamma}10^{-18}~ {\rm GeV^{-1} cm^{-2} s^{-1} sr^{-1}} \;,
\end{align}
which represents models consistent with the Fermi acceleration mechanism and extends to infinite energies.
In practice, the reachable energy of astrophysical accelerators always features a cutoff due to the Hillas criterion~\cite{Hillas:1984}.
For the cutoff model, the flux in eq.~(\ref{eq:dPhidE}) will be multiplied by a suppression factor  $\exp\left({-E^{}_{\nu}/E^{}_{\rm cutoff}}\right)$.
To confine the flux parameters, we construct a likelihood based on the results in ref.~\cite{IceCube:2020wum}:
\begin{align} \label{eq:}
-2\ln{\mathcal{L}^{}_{6\nu}} = \frac{({\rm \Phi^{}_{0}}- {\rm \Phi}^{\rm bf}_{0})^2}{ \sigma({\rm \Phi^{}_{0}})^2} + \frac{(\gamma- \gamma^{\rm bf})^2}{ \sigma(\gamma)^2} \;,
\end{align}
with the best-fit values ${\rm \Phi}^{\rm bf}_{0} = 6.37$ and $\gamma^{\rm bf} = 2.87$, as well as the $1\sigma$ errors $\sigma({\rm \Phi^{}_{0}}) = 1.54$ and $\sigma(\gamma) = 0.2$.
For the cutoff model we further derive the likelihood for $E^{}_{\rm cutoff}$ from fig.~ VI.9 of ref.~\cite{IceCube:2020wum} where the test-statistic has been marginalized.
Note that in this case we have ignored possible correlations among ${\rm \Phi}^{}_{0}$, $\gamma$ and $E^{}_{\rm cutoff}$, which are not provided. Nevertheless, such a choice will be more conservative because less information is utilized in our analysis.

\begin{figure}[t]
	\centering
	\includegraphics[width=0.6\textwidth]{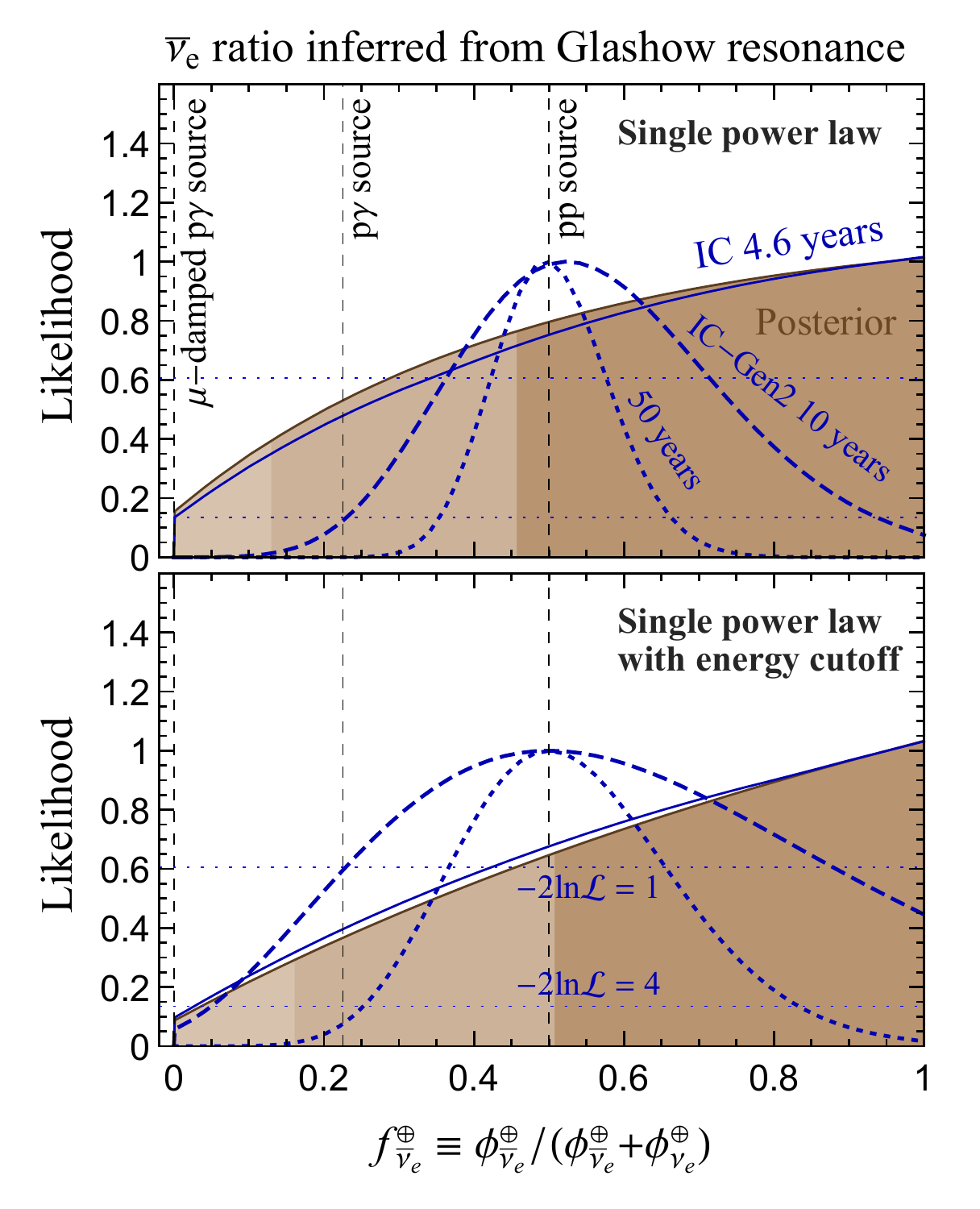}
	\caption{The likelihood (in blue) or posterior (in brown) of the $\overline{\nu}^{}_{e}$ fraction $f^{\oplus}_{\overline{\nu}_e} $ inferred from the Glashow resonance event in IceCube with $4.6$ years of data taking. The upper panel assumes a single power-law flux model with central values and uncertainties from ref.~\cite{IceCube:2020wum}, while the lower one has incorporated an exponential cutoff $E^{}_{\rm cutoff}$ in the neutrino spectrum. The expected $\overline{\nu}^{}_{e}$ fractions of three representative ultrahigh energy neutrino source models, including the ideal pp  ($f^{\oplus}_{\overline{\nu}_e} \approx 0.5$), the ideal p$\gamma$  ($f^{\oplus}_{\overline{\nu}_e} \approx 0.23$) and the ideal $\mu$-damped p$\gamma$ ($f^{\oplus}_{\overline{\nu}_e} \approx 0$) sources, are indicated by the dashed vertical lines. 
		When the p$\gamma$ collision energy at the source is very high, we may expect deviations from those ideal source models. By assuming an equal mixture of single-pion and multi-pion production at the source, we find the expected value $f^{\oplus}_{\overline{\nu}_e} \approx 0.36$ at Earth, shown as the dotted vertical lines. If multi-pion production is more dominant than this assumption, the expected $f^{\oplus}_{\overline{\nu}_e}$ should move to even larger values.
		The sensitivity of the future IceCube-Gen2 project with an effective exposure of ten (fifty) years is shown as the dashed (dotted) blue curves, assuming that the pp source is dominant with ${\rm \Phi}^{}_{0} = 6.37$, $\gamma = 2.7$ and $E^{}_{\rm cutoff} = 5~{\rm PeV}$.}
	\label{fig:Lnuebar}
\end{figure}

After the prior knowledge of $\{{\rm \Phi}^{}_{0}, \gamma(, E^{}_{\rm cutoff})\}$ has been established by HESE, we continue with fitting $f^{\oplus}_{\overline{\nu}_e}$ to PEPE. 
The task is to calculate the likelihood $\mathcal{L}^{}_{\overline{\nu}_{e}} (f^{\oplus}_{\overline{\nu}_e})$ with the  GR candidate we have.
The joint likelihood can then be obtained with $\mathcal{L}^{}_{\rm tot}=\mathcal{L}^{}_{6\nu} \times \mathcal{L}^{}_{\overline{\nu}_{e}}$ for the parameter set ${\rm \Theta} \equiv \{ {\rm \Phi}^{}_{0}, \gamma(, E^{}_{\rm cutoff}), f^{\oplus}_{\overline{\nu}_e}\}$.
In the frame of extended likelihood analysis of unbinned data~\cite{Cowan:1998ji},
the likelihood is calculated with
\begin{align} \label{eq:Lnuebar}
\mathcal{L}^{}_{\overline{\nu}_{e}} =  &  \prod_{i=1}^{n} \left[ \mu^{}_{\rm DIS}{P}^{}_{\rm DIS}(\# {i}| {\rm \Theta}) + \mu^{}_{\rm GR}{P}^{}_{\rm GR}(\# {i}| {\rm \Theta})\right] \times \frac{1}{n!} \mathrm{e}^{-(\mu^{}_{\rm DIS}+ \mu^{}_{\rm GR})}\;,
\end{align}
where $\mu^{}_{\rm DIS}$ and $\mu^{}_{\rm GR}$ are the expected event numbers within the energy window $E^{}_{\rm dep} \in [4, 10]~{\rm PeV}$ for the DIS and the GR, respectively, and $\# i$ represents in general all possible GR candidates. Moreover, ${P}^{}_{\rm DIS/GR}(\# {i}| {\rm \Theta})$ is the normalized probability to have an event at $\# {i}$'s energy for the given model parameter set ${\rm \Theta}$.
Since there is only one GR candidate so far we have $n = 1$ in eq.~(\ref{eq:Lnuebar}).
The event numbers can be obtained by integrating the flux and cross sections with the detector configuration.

\section{Main results}%
With the framework  above, we can compute the total likelihood $\mathcal{L}^{}_{\rm tot}$ as a function of the parameter set $\{ {\rm \Phi}^{}_{0}, \gamma(, E^{}_{\rm cutoff}), f^{\oplus}_{\overline{\nu}_e}\}$. 
The likelihood can then be used for either frequentist or Bayesian interpretations. 
For the frequentist interpretation, we 
obtain the likelihood maximum $\mathcal{L}^{\rm max}_{\rm tot} (f^{\oplus}_{\overline{\nu}_e})$
by marginalizing over the other parameters.
For the Bayesian interpretation, we need to derive the posterior distribution of $f^{\oplus}_{\overline{\nu}_e}$ by integrating over  the likelihood and priors. We choose flat priors on ${\rm \Phi}^{}_{0}$, $\gamma$, $f^{\oplus}_{\overline{\nu}_e}$ and $\ln{E^{}_{\rm cutoff}}$ for illustration.

Our main results are given in fig.~\ref{fig:Lnuebar}, which shows the likelihood function (in blue) or posterior distribution (in brown) of the $\overline{\nu}^{}_{e}$ fraction $f^{\oplus}_{\overline{\nu}_{e}} $ inferred from the IceCube 4.6-year data. Uncertainties from neutrino flux parameters have been systematically included and marginalized when we constrain $f^{\oplus}_{\overline{\nu}_{e}}$. 
The upper and lower panels stand for the assumptions of an unbroken single power-law flux model and a single power-law model with a varying exponential energy cutoff, respectively~\cite{IceCube:2020wum}.
For blue curves, the horizontal lines with $-2\ln{\mathcal{L}} = 1$ and $4$ roughly set the $1\sigma$ and $2\sigma$ confidence levels. For brown regions, the $1\sigma$ and $2\sigma$ credible intervals have been covered from dark to light colors.

We find that for all cases, the $\mu$-damped p$\gamma$ source with $f^{\oplus}_{\overline{\nu}_{e}} \approx 0$ (single-pion production via the $\Delta$-resonance for the ideal scenario) is  excluded by around $2\sigma$ level. The current IceCube 4.6-year data weakly favor the pp source but are not able to exclude the ideal p$\gamma$ source considerably (only at $1\sigma$ or so); see the dashed vertical lines. 
While interpreting the above results, one must keep in mind that neutrinos may not only be produced by the ideal $\Delta$-resonance of the p$\gamma$ scattering, but also by other possible effects that can dominate at high energies~\cite{Baerwald_2011, Huemmer_2010}, such as multi-pion production, higher resonances, and the direct (t-channel) production of pions.
Note that the above considerations do not affect our model-independent results of $f^{\oplus}_{\overline{\nu}_e}$ extracted from experimental data.
For those cases, the theoretically expected value of $f^{\oplus}_{\overline{\nu}_e}$ for the $p\gamma$ source will  shift towards larger values. The actual magnitude of the deviation  depends on the details of the $\pi^+$ and $\pi^-$ mixture at the source. 
For demonstration, we assume that the single-pion and multi-pion channels have the same production rate at the source and draw the expected value $f^{\oplus}_{\overline{\nu}_e} \approx 0.36$ as the dotted vertical lines in fig.~\ref{fig:Lnuebar}.
If the multi-pion channel contributes more, this vertical line should move even further to the right.
On the other hand, for the pp source the multi-pion contribution does not change the expected value of $f^{\oplus}_{\overline{\nu}_e}$.

\begin{figure}[t!]
	\centering 
 \includegraphics[width=0.6\textwidth]{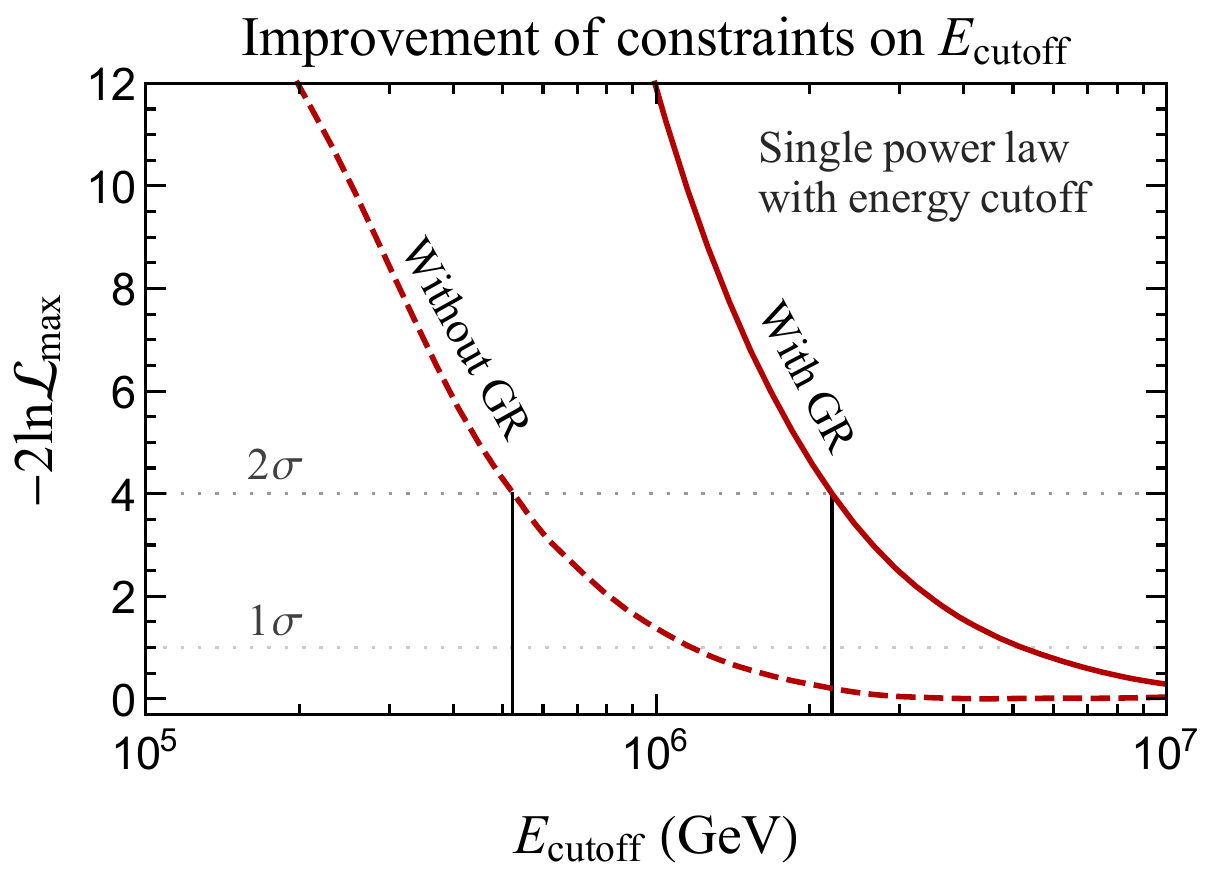}
	\caption{The log-likelihood of the energy cutoff $E^{}_{\rm cutoff}$. The dashed curve is taken from fig.~ VI.9 of ref.~\cite{IceCube:2020wum}, while the solid curve is derived from the Glashow resonance candidate event by marginalizing over the other model parameters. }
	\label{fig:lhEcut}
\end{figure}

Last but not least we should emphasize that the GR event can also constrain the possible energy cutoff $E^{}_{\rm cutoff}$ in the neutrino spectrum. The original best-fit value of $E^{}_{\rm cutoff}$ without GR is around $5~{\rm PeV}$ in ref.~\cite{IceCube:2020wum}, with a $2\sigma$ lower boundary at $0.5~{\rm PeV}$. The presence of the GR candidate event will push the $2\sigma$ lower boundary to $2.2~{\rm PeV}$, as illustrated in fig.~\ref{fig:lhEcut}.

\section{Outlook}%
Using the recent GR candidate event identified by IceCube, we have performed an analysis to infer the $\overline{\nu}^{}_{e}$ content in UHE astrophysical neutrinos.
We treat the $\overline{\nu}^{}_{e}$ fraction as a model-independent free parameter and have set a generic constraint on it by including the uncertainties in the UHE neutrino flux. From the candidate event measured so far, we find a weak preference for the pp source under the ideal assumption. The situation will be greatly improved by the upcoming next-generation neutrino telescopes.
If the neutrino production in the p$\gamma$ source is dominated by other channels at higher energies such as multi-pion production, we would need additional information from the multi-wavelength observations of the source to distinguish between pp and p$\gamma$ sources.


In the future, there are many projects such as IceCube-Gen2~\cite{IceCube:2014gqr,IceCube-Gen2:2020qha}, Baikal-GVD~\cite{Baikal-GVD:2018isr}, KM3NeT~\cite{KM3Net:2016zxf}, P-ONE~\cite{P-ONE:2020ljt}, TAMBO~\cite{Romero-Wolf:2020pzh}, TRIDENT~\cite{Ye:2022vbk} and so on, which will provide very valuable sensitivities to PeV astrophysical neutrinos~\cite{Huang:2021mki,Coleman:2022abf,Ackermann:2022rqc,Valera:2022wmu}. 
We take IceCube-Gen2 for demonstration by rescaling the current IceCube target mass by ten times, and perform a count analysis in the energy window of $[4, 10]~{\rm PeV}$.
The sensitivity for ten (fifty) years of effective exposure is shown as the dashed (dotted) curves in fig.~\ref{fig:Lnuebar}. Because the flux parameters $\{ {\rm \Phi}^{}_{0}, \gamma(, E^{}_{\rm cutoff})\}$  can be very precisely determined in the future~\cite{IceCube-Gen2:2020qha}, we choose a reasonably optimistic spectrum as ${\rm \Phi}^{}_{0} = 6.37$, $\gamma = 2.7$ and $E^{}_{\rm cutoff} = 5~{\rm PeV}$ in making the forecast; see fig.~16 of ref.~\cite{IceCube-Gen2:2020qha} for example. 
It is worth noting that the tau neutrino telescope TAMBO can be sensitive to the Glashow resonance event by searching for the tau-induced showers from $W$ decays. Unfortunately, with a dedicated simulation close to the TAMBO setup, we find that the event number is too small, e.g., only two events with an optimistic flux and ten years of exposure, compared to the DIS background of $\mathcal{O}(100)$.
We will elaborate on the related analysis in a future work.

Assuming the pp type as the true source, i.e., $f^{\oplus}_{\overline{\nu}_{e}}=0.5$, we expect eleven GR events in IceCube-Gen2 with ten years of exposure for the optimistic single power-law model. If we take an exponential cutoff $E^{}_{\rm cutoff} = 5~{\rm PeV}$ in the spectrum, the event expectation would be reduced to three. The expected number of events is still diverse due to low statistics of events at PeV energies. 
For the single power-law model, IceCube-Gen2 with ten years of exposure can already differentiate ideal pp from p$\gamma$ sources with a $2\sigma$ confidence level. However, if there is an exponential cutoff at $5$ PeV, an effective exposure of fifty years would be required to reach the $2\sigma$ level.
Those results can also be applied to other telescopes by adjusting the effective exposure.
By measuring the spectrum precisely in the future, one may go beyond the assumptions of single power-law flux model (with cutoff) and take the spectrum with a general energy dependence.

The hybrid cascade and early muon reconstruction in IceCube can already greatly improve the angular resolution of the GR shower.
In case of the increased statistics, GR events detected in future experiments can also be used to produce a map of the sky and identify associated PeVatrons~\cite{	LHAASO:2021cbz,LHAASO_nature,Sudoh:2022sdk}. Our main point is that knowledge about neutrino sources will be significantly improved by those upcoming facilities with large statistics, which also guarantees a robust frontier for possible new physics studies~\cite{Bustamante:2020niz,Jezo:2014kla,Babu:2019vff,Dey:2020fbx,Babu:2022fje,Xu:2022svm,Arguelles:2022xxa,Huang:2022pce,Huang:2022ebg,Heighton:2023qpg}.

\appendix
\section{Appendix: Details of the Doppler broadening effect}\label{AppendixDoppler}
\begin{figure}[h!]
	\centering
	\includegraphics[width=0.7\columnwidth]{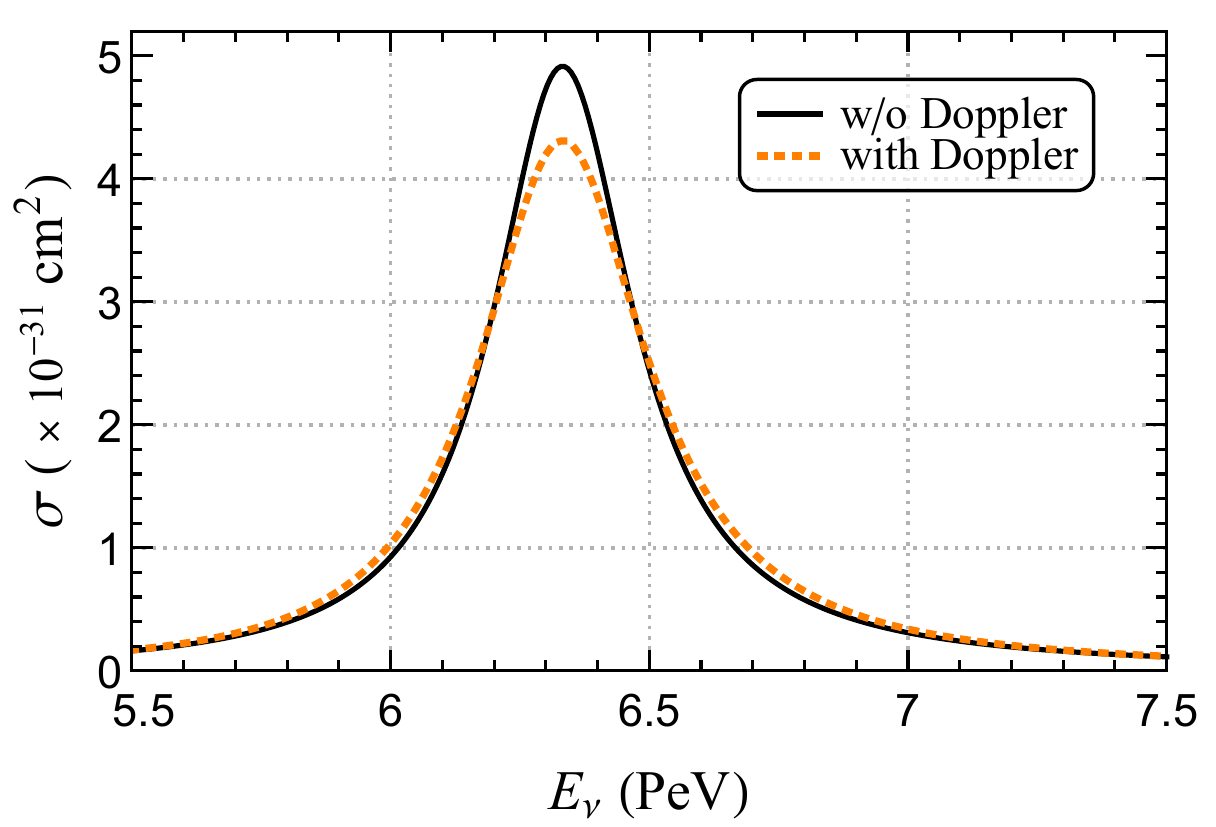}
	\caption{Cross section for the Glashow resonance process $\Bar{\nu}_e+e^- \rightarrow W^- \rightarrow X$ with and without Doppler broadening and assuming ice (${\rm H}^{}_{2} {\rm O}$) as the target. The black curve represents the cross section without Doppler broadening and for the orange curve Doppler broadening is included.}
	\label{fig:DopplerBroadened}
\end{figure}
\noindent
We  follow the procedure outlined in ref.~\cite{Glashow:2014} to include the Doppler broadening effect of atomic electrons. By integrating over angular variables in eq.~(\ref{eq:integrationDoppler}), we  arrive at
\begin{align}
\label{eq:sigmaorbitalintegrated}
\sigma (E^{}_{\nu})  = \frac{6\pi {\rm \Gamma}_W^2 {\rm Br}^{}_{W^-\rightarrow\overline{\nu}_e e^-}}{M^{}_W m^{}_e E^{}_{\nu} }\int & d\beta \frac{F(\beta)}{\beta} 
\left\{\frac{1}{2M^{}_W}\left[ \ln (y_h^2+1)-\ln (y_l^2+1) \right] \right. \notag
\\ &  \left. +  \frac{1}{{\rm \Gamma}^{}_W}\left[ \arctan (y^{}_h)-\arctan(y^{}_l) \right] \right\}
\end{align}
where
\begin{equation}
y^{}_h=\frac{2m^{}_eE^{}_{\nu}(1+\beta )+m_e^2-M_W^2}{{\rm \Gamma}^{}_W M^{}_W}
\text{\hspace{0.5cm} and  \hspace{0.5cm}}
y^{}_l=\frac{2 m^{}_e E^{}_{\nu} (1-\beta ) + m_e^2 - M_W^2}{{\rm \Gamma}^{}_W M^{}_{W}}\,.
\end{equation}
Now the problem is attributed to the integration over the averaged electron velocity distribution $F(\beta)$.
In terms of the wave function of an electron with quantum numbers $n$ and $l$, the distribution reads
\begin{equation}
f^{}_{n\,l}(\beta)=m_e\int \mathrm{d}{\rm \Omega}^{}_k k^2 |{\rm \Psi}^{}_{n\,l}(k)|^2
\text{\hspace{0.5cm} with  \hspace{0.5cm}}
{\rm \Psi}^{}_{n\,l}(\mathbf{k})\propto Y_{l\,m}^*({\rm \Omega}^{}_k)\int_0^{\infty} \mathrm{d}r \, r^{n+1}e^{-\mu r} j^{}_l(kr)\;,
\end{equation}
where $k=m^{}_e \beta$ and $\mu^{}_{n\,l}=\xi^{}_{n\,l}/a^{}_0$. Here, $a_0$ denotes the Bohr radius, $\xi^{}_{n\,l}=Z^{}_{\rm eff}/n=(Z-\sigma^{}_{n\,l})/n$, and $\sigma_{n\,l}$ accounts for the screening of the nuclear charge by the other electrons in the atom.

After the integration, we can get the velocity distribution for atoms up to $Z=26$~\cite{Glashow:2014}:
\begin{align}
f^{}_{1s}(k)=&\frac{32}{\pi}\frac{\mu_{1s}^5k^2}{(\mu_{1s}^2+k^2)^4} \;,\\
f^{}_{2s}(k)=&\frac{32}{3\pi}\frac{\mu_{2s}^5(3\mu_{2s}^2k-k^3)^2}{(\mu_{2s}^2+k^2)^6} \;, \label{eq:correct1}\\
f^{}_{2p}(k)=&\frac{512}{3\pi}\frac{\mu_{2p}^7k^4}{(\mu_{2p}^2+k^2)^6} \;, \\
f^{}_{3s}(k)=&\frac{1024}{5\pi}\frac{\mu_{3s}^7 (\mu^3_{3s} k -\mu^{}_{3s} k^3)^2}{(\mu_{3s}^2+k^2)^8} \;, \\
f^{}_{3p}(k)=&\frac{1024}{45\pi}\frac{\mu_{3p}^7 (5\mu^2_{3p} k^2 - k^4)^2}{(\mu_{3p}^2+k^2)^8} \;,\\
f^{}_{3d}(k)=&\frac{4096}{5\pi}\frac{\mu_{3d}^9 k^6}{(\mu_{3d}^2+k^2)^8} \;, \\
f^{}_{4s}(k)=&\frac{512}{35\pi}\frac{\mu_{4s}^9 (5\mu^4_{4s}k -10\mu^2_{4s} k^3 + k^5)^2}{(\mu_{4s}^2+k^2)^{10}} \;.\label{eq:correct2}
\end{align}
Note that we have checked the expressions in ref.~\cite{Glashow:2014} and corrected possible discrepancies as in our eqs.~(\ref{eq:correct1}) and (\ref{eq:correct2}).

We take the ice molecule ${\rm H}^{}_{2} {\rm O}$ as an example. For oxygen, $\mu_{1s}=7.6579$, $\mu_{2s}=2.2458$ and $\mu_{2p}=2.2266$~\cite{Clementi:1963}, and for hydrogen $\mu_{1s}=1$.
We weigh the distribution functions by averaging over the electron numbers:
\begin{equation}
\label{eq:Fice}
F^{}_{\rm ice}(\beta)= \frac{2F^{}_{\rm H}(\beta )+8F^{}_{\rm O}(\beta )}{10}\;.
\end{equation}
Using eq.~(\ref{eq:sigmaorbitalintegrated}) together with eq.~(\ref{eq:Fice}) we get the Doppler broadened cross section for ice as the target, which is depicted in fig.~\ref{fig:DopplerBroadened}.
The effect reduces the peak by about $12\,\%$. Even though the total cross section integrated over the initial neutrino energy is barely altered, the broadening effect will make a difference when a non-uniform neutrino spectrum is considered.

\acknowledgments
GYH would like to thank Xiao-Jun Bi, Xun-Jie Xu, Shoushan Zhang and Shun Zhou for inspiring comments and discussions. GYH is supported in part by the Alexander von Humboldt Foundation.

\bibliographystyle{utcaps_mod}
\bibliography{reference}

\providecommand{\href}[2]{#2}\begingroup\raggedright\begin{thebibliography}{10}

\bibitem{IceCube:2021rpz}
{\normalfont \bfseries IceCube}, M.~G. Aartsen {\em et al.}, ``{\em {Detection
  of a particle shower at the Glashow resonance with IceCube}},''
  \href{http://dx.doi.org/10.1038/s41586-021-03256-1}{Nature {\normalfont
  \bfseries 591} (2021) no.~7849, 220--224},
  \href{http://arxiv.org/abs/2110.15051}{{\normalfont \ttfamily
  arXiv:2110.15051}}. [Erratum: Nature 592, E11 (2021)].

\bibitem{IceCube:2013low}
{\normalfont \bfseries IceCube}, M.~G. Aartsen {\em et al.}, ``{\em {Evidence
  for High-Energy Extraterrestrial Neutrinos at the IceCube Detector}},''
  \href{http://dx.doi.org/10.1126/science.1242856}{Science {\normalfont
  \bfseries 342} (2013)  1242856},
  \href{http://arxiv.org/abs/1311.5238}{{\normalfont \ttfamily
  arXiv:1311.5238}}.

\bibitem{IceCube:2013cdw}
{\normalfont \bfseries IceCube}, M.~G. Aartsen {\em et al.}, ``{\em {First
  observation of PeV-energy neutrinos with IceCube}},''
  \href{http://dx.doi.org/10.1103/PhysRevLett.111.021103}{Phys. Rev. Lett.
  {\normalfont \bfseries 111} (2013)  021103},
  \href{http://arxiv.org/abs/1304.5356}{{\normalfont \ttfamily
  arXiv:1304.5356}}.

\bibitem{IceCube:2018cha}
{\normalfont \bfseries IceCube}, M.~G. Aartsen {\em et al.}, ``{\em {Neutrino
  emission from the direction of the blazar TXS 0506+056 prior to the
  IceCube-170922A alert}},''
  \href{http://dx.doi.org/10.1126/science.aat2890}{Science {\normalfont
  \bfseries 361} (2018) no.~6398, 147--151},
  \href{http://arxiv.org/abs/1807.08794}{{\normalfont \ttfamily
  arXiv:1807.08794}}.

\bibitem{IceCube:2020abv}
{\normalfont \bfseries IceCube}, R.~Abbasi {\em et al.}, ``{\em {Measurement of
  Astrophysical Tau Neutrinos in IceCube's High-Energy Starting Events}},''
  \href{http://arxiv.org/abs/2011.03561}{{\normalfont \ttfamily
  arXiv:2011.03561}}.

\bibitem{IceCube:2020wum}
{\normalfont \bfseries IceCube}, R.~Abbasi {\em et al.}, ``{\em {The IceCube
  high-energy starting event sample: Description and flux characterization with
  7.5 years of data}},''
  \href{http://dx.doi.org/10.1103/PhysRevD.104.022002}{Phys. Rev. D
  {\normalfont \bfseries 104} (2021)  022002},
  \href{http://arxiv.org/abs/2011.03545}{{\normalfont \ttfamily
  arXiv:2011.03545}}.

\bibitem{IceCube:2022der}
{\normalfont \bfseries IceCube}, R.~Abbasi {\em et al.}, ``{\em {Evidence for
  neutrino emission from the nearby active galaxy NGC 1068}},''
  \href{http://dx.doi.org/10.1126/science.abg3395}{Science {\normalfont
  \bfseries 378} (2022) no.~6619, 538--543},
  \href{http://arxiv.org/abs/2211.09972}{{\normalfont \ttfamily
  arXiv:2211.09972}}.

\bibitem{Halzen:2022pez}
F.~Halzen and A.~Kheirandish, {\em {Chapter 5: IceCube and High-Energy Cosmic
  Neutrinos}},
  \href{http://dx.doi.org/10.1142/9789811282645_0005}{pp.~107--235}.
\newblock 2023.
\newblock \href{http://arxiv.org/abs/2202.00694}{{\normalfont \ttfamily
  arXiv:2202.00694}}.

\bibitem{Gaisser:1994yf}
T.~K. Gaisser, F.~Halzen, and T.~Stanev, ``{\em {Particle astrophysics with
  high-energy neutrinos}},''
  \href{http://dx.doi.org/10.1016/0370-1573(95)00003-Y}{Phys. Rept.
  {\normalfont \bfseries 258} (1995)  173--236},
  \href{http://arxiv.org/abs/hep-ph/9410384}{{\normalfont \ttfamily
  arXiv:hep-ph/9410384}}. [Erratum: Phys.Rept. 271, 355--356 (1996)].

\bibitem{Bhattacharjee:1999mup}
P.~Bhattacharjee and G.~Sigl, ``{\em {Origin and propagation of extremely
  high-energy cosmic rays}},''
  \href{http://dx.doi.org/10.1016/S0370-1573(99)00101-5}{Phys. Rept.
  {\normalfont \bfseries 327} (2000)  109--247},
  \href{http://arxiv.org/abs/astro-ph/9811011}{{\normalfont \ttfamily
  arXiv:astro-ph/9811011}}.

\bibitem{Beatty:2009zz}
J.~J. Beatty and S.~Westerhoff, ``{\em {The Highest-Energy Cosmic Rays}},''
  \href{http://dx.doi.org/10.1146/annurev.nucl.58.110707.171154}{Ann. Rev.
  Nucl. Part. Sci. {\normalfont \bfseries 59} (2009)  319--345}.

\bibitem{gaisser_engel_resconi_2016}
T.~K. Gaisser, R.~Engel, and E.~Resconi,
  \href{http://dx.doi.org/10.1017/CBO9781139192194}{{\em Cosmic Rays and
  Particle Physics}}.
\newblock Cambridge University Press, 2~ed., 2016.

\bibitem{anchordoqui2019ultra}
L.~A. Anchordoqui, ``{\em Ultra-high-energy cosmic rays},'' Physics Reports
  {\normalfont \bfseries 801} (2019)  1--93.

\bibitem{tjus2020closing}
J.~B. Tjus and L.~Merten, ``{\em Closing in on the origin of Galactic cosmic
  rays using multimessenger information},'' Physics Reports {\normalfont
  \bfseries 872} (2020)  1--98.

\bibitem{Murase:2019tjj}
K.~Murase and I.~Bartos, ``{\em {High-Energy Multimessenger Transient
  Astrophysics}},''
  \href{http://dx.doi.org/10.1146/annurev-nucl-101918-023510}{Ann. Rev. Nucl.
  Part. Sci. {\normalfont \bfseries 69} (2019)  477--506},
  \href{http://arxiv.org/abs/1907.12506}{{\normalfont \ttfamily
  arXiv:1907.12506}}.

\bibitem{Murase:2022feu}
K.~Murase and F.~W. Stecker, ``{\em {High-Energy Neutrinos from Active Galactic
  Nuclei}},'' \href{http://arxiv.org/abs/2202.03381}{{\normalfont \ttfamily
  arXiv:2202.03381}}.

\bibitem{Troitsky:2021nvu}
S.~Troitsky, ``{\em {Constraints on models of the origin of high-energy
  astrophysical neutrinos}},''
  \href{http://dx.doi.org/10.3367/UFNe.2021.09.039062}{Usp. Fiz. Nauk
  {\normalfont \bfseries 191} (2021) no.~12, 1333--1360},
  \href{http://arxiv.org/abs/2112.09611}{{\normalfont \ttfamily
  arXiv:2112.09611}}.

\bibitem{Xing:2011zza}
Z.-z. Xing and S.~Zhou, {\em {Neutrinos in particle physics, astronomy and
  cosmology}}.
\newblock Zhejiang University Press, 2011.

\bibitem{Mena:2014sja}
O.~Mena, S.~Palomares-Ruiz, and A.~C. Vincent, ``{\em {Flavor Composition of
  the High-Energy Neutrino Events in IceCube}},''
  \href{http://dx.doi.org/10.1103/PhysRevLett.113.091103}{Phys. Rev. Lett.
  {\normalfont \bfseries 113} (2014)  091103},
\href{http://arxiv.org/abs/1404.0017}{{\normalfont \ttfamily arXiv:1404.0017}}.

\bibitem{Chen:2014gxa}
C.-Y. Chen, P.~S. Bhupal~Dev, and A.~Soni, ``{\em {Two-component flux
  explanation for the high energy neutrino events at IceCube}},''
  \href{http://dx.doi.org/10.1103/PhysRevD.92.073001}{Phys. Rev. {\normalfont
  \bfseries D92} (2015) no.~7, 073001},
\href{http://arxiv.org/abs/1411.5658}{{\normalfont \ttfamily arXiv:1411.5658}}.

\bibitem{Palomares-Ruiz:2015mka}
S.~Palomares-Ruiz, A.~C. Vincent, and O.~Mena, ``{\em {Spectral analysis of the
  high-energy IceCube neutrinos}},''
  \href{http://dx.doi.org/10.1103/PhysRevD.91.103008}{Phys. Rev. {\normalfont
  \bfseries D91} (2015) no.~10, 103008},
\href{http://arxiv.org/abs/1502.02649}{{\normalfont \ttfamily
  arXiv:1502.02649}}.

\bibitem{Aartsen:2015ivb}
{\normalfont \bfseries IceCube}, M.~G. Aartsen {\em et al.}, ``{\em {Flavor
  Ratio of Astrophysical Neutrinos above 35 TeV in IceCube}},''
  \href{http://dx.doi.org/10.1103/PhysRevLett.114.171102}{Phys. Rev. Lett.
  {\normalfont \bfseries 114} (2015) no.~17, 171102},
\href{http://arxiv.org/abs/1502.03376}{{\normalfont \ttfamily
  arXiv:1502.03376}}.

\bibitem{Palladino:2015zua}
A.~Palladino, G.~Pagliaroli, F.~L. Villante, and F.~Vissani, ``{\em {What is
  the Flavor of the Cosmic Neutrinos Seen by IceCube?}},''
  \href{http://dx.doi.org/10.1103/PhysRevLett.114.171101}{Phys. Rev. Lett.
  {\normalfont \bfseries 114} (2015) no.~17, 171101},
\href{http://arxiv.org/abs/1502.02923}{{\normalfont \ttfamily
  arXiv:1502.02923}}.

\bibitem{Arguelles:2015dca}
C.~A. Argüelles, T.~Katori, and J.~Salvado, ``{\em {New Physics in
  Astrophysical Neutrino Flavor}},''
  \href{http://dx.doi.org/10.1103/PhysRevLett.115.161303}{Phys. Rev. Lett.
  {\normalfont \bfseries 115} (2015)  161303},
\href{http://arxiv.org/abs/1506.02043}{{\normalfont \ttfamily
  arXiv:1506.02043}}.

\bibitem{Bustamante:2015waa}
M.~Bustamante, J.~F. Beacom, and W.~Winter, ``{\em {Theoretically palatable
  flavor combinations of astrophysical neutrinos}},''
  \href{http://dx.doi.org/10.1103/PhysRevLett.115.161302}{Phys. Rev. Lett.
  {\normalfont \bfseries 115} (2015) no.~16, 161302},
  \href{http://arxiv.org/abs/1506.02645}{{\normalfont \ttfamily
  arXiv:1506.02645}}.

\bibitem{Aartsen:2015knd}
{\normalfont \bfseries IceCube}, M.~G. Aartsen {\em et al.}, ``{\em {A combined
  maximum-likelihood analysis of the high-energy astrophysical neutrino flux
  measured with IceCube}},''
  \href{http://dx.doi.org/10.1088/0004-637X/809/1/98}{Astrophys. J.
  {\normalfont \bfseries 809} (2015) no.~1, 98},
\href{http://arxiv.org/abs/1507.03991}{{\normalfont \ttfamily
  arXiv:1507.03991}}.

\bibitem{Brdar:2016thq}
V.~Brdar, J.~Kopp, and X.-P. Wang, ``{\em {Sterile Neutrinos and Flavor Ratios
  in IceCube}},'' \href{http://dx.doi.org/10.1088/1475-7516/2017/01/026}{JCAP
  {\normalfont \bfseries 01} (2017)  026},
  \href{http://arxiv.org/abs/1611.04598}{{\normalfont \ttfamily
  arXiv:1611.04598}}.

\bibitem{DAmico:2017dwq}
G.~D'Amico, ``{\em {Flavor and energy inference for the high-energy IceCube
  neutrinos}},''
  \href{http://dx.doi.org/10.1016/j.astropartphys.2018.04.002}{Astropart. Phys.
  {\normalfont \bfseries 101} (2018)  8--16},
  \href{http://arxiv.org/abs/1712.04979}{{\normalfont \ttfamily
  arXiv:1712.04979}}.

\bibitem{Pagliaroli:2015rca}
G.~Pagliaroli, A.~Palladino, F.~L. Villante, and F.~Vissani, ``{\em {Testing
  nonradiative neutrino decay scenarios with IceCube data}},''
  \href{http://dx.doi.org/10.1103/PhysRevD.92.113008}{Phys. Rev. D {\normalfont
  \bfseries 92} (2015) no.~11, 113008},
  \href{http://arxiv.org/abs/1506.02624}{{\normalfont \ttfamily
  arXiv:1506.02624}}.

\bibitem{Rasmussen:2017ert}
R.~W. Rasmussen, L.~Lechner, M.~Ackermann, M.~Kowalski, and W.~Winter, ``{\em
  {Astrophysical neutrinos flavored with Beyond the Standard Model physics}},''
  \href{http://dx.doi.org/10.1103/PhysRevD.96.083018}{Phys. Rev. D {\normalfont
  \bfseries 96} (2017) no.~8, 083018},
  \href{http://arxiv.org/abs/1707.07684}{{\normalfont \ttfamily
  arXiv:1707.07684}}.

\bibitem{Brdar:2018tce}
V.~Brdar and R.~S.~L. Hansen, ``{\em {IceCube Flavor Ratios with Identified
  Astrophysical Sources: Towards Improving New Physics Testability}},''
  \href{http://dx.doi.org/10.1088/1475-7516/2019/02/023}{JCAP {\normalfont
  \bfseries 02} (2019)  023},
  \href{http://arxiv.org/abs/1812.05541}{{\normalfont \ttfamily
  arXiv:1812.05541}}.

\bibitem{Bustamante:2019sdb}
M.~Bustamante and M.~Ahlers, ``{\em {Inferring the flavor of high-energy
  astrophysical neutrinos at their sources}},''
  \href{http://dx.doi.org/10.1103/PhysRevLett.122.241101}{Phys. Rev. Lett.
  {\normalfont \bfseries 122} (2019) no.~24, 241101},
\href{http://arxiv.org/abs/1901.10087}{{\normalfont \ttfamily
  arXiv:1901.10087}}.

\bibitem{Palladino:2019pid}
A.~Palladino, ``{\em {The flavor composition of astrophysical neutrinos after 8
  years of IceCube: an indication of neutron decay scenario?}},''
  \href{http://dx.doi.org/10.1140/epjc/s10052-019-7018-7}{Eur. Phys. J.
  {\normalfont \bfseries C79} (2019) no.~6, 500},
\href{http://arxiv.org/abs/1902.08630}{{\normalfont \ttfamily
  arXiv:1902.08630}}.

\bibitem{Stachurska:2019srh}
{\normalfont \bfseries IceCube}, J.~Stachurska, ``{\em {IceCube High Energy
  Starting Events at 7.5 Years -- New Measurements of Flux and Flavor}},''
  \href{http://dx.doi.org/10.1051/epjconf/201920702005}{EPJ Web Conf.
  {\normalfont \bfseries 207} (2019)  02005},
\href{http://arxiv.org/abs/1905.04237}{{\normalfont \ttfamily
  arXiv:1905.04237}}.

\bibitem{Song:2020nfh}
N.~Song, S.~W. Li, C.~A. Arg\"uelles, M.~Bustamante, and A.~C. Vincent, ``{\em
  {The Future of High-Energy Astrophysical Neutrino Flavor Measurements}},''
  \href{http://dx.doi.org/10.1088/1475-7516/2021/04/054}{JCAP {\normalfont
  \bfseries 04} (2021)  054},
  \href{http://arxiv.org/abs/2012.12893}{{\normalfont \ttfamily
  arXiv:2012.12893}}.

\bibitem{Glashow:1960zz}
S.~L. Glashow, ``{\em {Resonant Scattering of Antineutrinos}},''
\href{http://dx.doi.org/10.1103/PhysRev.118.316}{Phys. Rev. {\normalfont
  \bfseries 118} (1960)  316--317}.

\bibitem{Berezinsky:1977sf}
V.~S. Berezinsky and A.~Z. Gazizov, ``{\em {Cosmic neutrino and the possibility
  of Searching for W bosons with masses 30-100 GeV in underwater
  experiments}},'' JETP Lett. {\normalfont \bfseries 25} (1977)  254--256.

\bibitem{Brown:1981ns}
R.~W. Brown and F.~W. Stecker, ``{\em {Cosmic Ray Neutrino Tests for Heavier
  Weak Bosons and Cosmic Antimatter}},''
  \href{http://dx.doi.org/10.1103/PhysRevD.26.373}{Phys. Rev. D {\normalfont
  \bfseries 26} (1982)  373}.

\bibitem{Anchordoqui:2004eb}
L.~A. Anchordoqui, H.~Goldberg, F.~Halzen, and T.~J. Weiler, ``{\em {Neutrinos
  as a diagnostic of high energy astrophysical processes}},''
  \href{http://dx.doi.org/10.1016/j.physletb.2005.06.056}{Phys. Lett.
  {\normalfont \bfseries B621} (2005)  18--21},
\href{http://arxiv.org/abs/hep-ph/0410003}{{\normalfont \ttfamily
  arXiv:hep-ph/0410003}}.

\bibitem{Hummer:2010ai}
S.~Hummer, M.~Maltoni, W.~Winter, and C.~Yaguna, ``{\em {Energy dependent
  neutrino flavor ratios from cosmic accelerators on the Hillas plot}},''
  \href{http://dx.doi.org/10.1016/j.astropartphys.2010.07.003}{Astropart. Phys.
  {\normalfont \bfseries 34} (2010)  205--224},
\href{http://arxiv.org/abs/1007.0006}{{\normalfont \ttfamily arXiv:1007.0006}}.

\bibitem{Xing:2011zm}
Z.-z. Xing and S.~Zhou, ``{\em {The Glashow resonance as a discriminator of UHE
  cosmic neutrinos originating from p-gamma and p-p collisions}},''
  \href{http://dx.doi.org/10.1103/PhysRevD.84.033006}{Phys. Rev. {\normalfont
  \bfseries D84} (2011)  033006},
\href{http://arxiv.org/abs/1105.4114}{{\normalfont \ttfamily arXiv:1105.4114}}.

\bibitem{Bhattacharya:2011qu}
A.~Bhattacharya, R.~Gandhi, W.~Rodejohann, and A.~Watanabe, ``{\em {The Glashow
  resonance at IceCube: signatures, event rates and $pp$ vs. $p\gamma$
  interactions}},'' \href{http://dx.doi.org/10.1088/1475-7516/2011/10/017}{JCAP
  {\normalfont \bfseries 1110} (2011)  017},
\href{http://arxiv.org/abs/1108.3163}{{\normalfont \ttfamily arXiv:1108.3163}}.

\bibitem{Bhattacharya:2012fh}
A.~Bhattacharya, R.~Gandhi, W.~Rodejohann, and A.~Watanabe, ``{\em {On the
  interpretation of IceCube cascade events in terms of the Glashow
  resonance}},''
\href{http://arxiv.org/abs/1209.2422}{{\normalfont \ttfamily arXiv:1209.2422}}.

\bibitem{Barger:2012mz}
V.~Barger, J.~Learned, and S.~Pakvasa, ``{\em {IceCube PeV Cascade Events
  Initiated by Electron-Antineutrinos at Glashow Resonance}},''
  \href{http://dx.doi.org/10.1103/PhysRevD.87.037302}{Phys. Rev. {\normalfont
  \bfseries D87} (2013) no.~3, 037302},
\href{http://arxiv.org/abs/1207.4571}{{\normalfont \ttfamily arXiv:1207.4571}}.

\bibitem{Barger:2014iua}
V.~Barger, L.~Fu, J.~G. Learned, D.~Marfatia, S.~Pakvasa, and T.~J. Weiler,
  ``{\em {Glashow resonance as a window into cosmic neutrino sources}},''
  \href{http://dx.doi.org/10.1103/PhysRevD.90.121301}{Phys. Rev. {\normalfont
  \bfseries D90} (2014)  121301},
\href{http://arxiv.org/abs/1407.3255}{{\normalfont \ttfamily arXiv:1407.3255}}.

\bibitem{Palladino:2015uoa}
A.~Palladino, G.~Pagliaroli, F.~L. Villante, and F.~Vissani, ``{\em {Double
  pulses and cascades above 2 PeV in IceCube}},''
  \href{http://dx.doi.org/10.1140/epjc/s10052-016-3893-3}{Eur. Phys. J.
  {\normalfont \bfseries C76} (2016) no.~2, 52},
\href{http://arxiv.org/abs/1510.05921}{{\normalfont \ttfamily
  arXiv:1510.05921}}.

\bibitem{Shoemaker:2015qul}
I.~M. Shoemaker and K.~Murase, ``{\em {Probing BSM Neutrino Physics with Flavor
  and Spectral Distortions: Prospects for Future High-Energy Neutrino
  Telescopes}},'' \href{http://dx.doi.org/10.1103/PhysRevD.93.085004}{Phys.
  Rev. {\normalfont \bfseries D93} (2016) no.~8, 085004},
\href{http://arxiv.org/abs/1512.07228}{{\normalfont \ttfamily
  arXiv:1512.07228}}.

\bibitem{Anchordoqui:2016ewn}
L.~A. Anchordoqui, M.~M. Block, L.~Durand, P.~Ha, J.~F. Soriano, and T.~J.
  Weiler, ``{\em {Evidence for a break in the spectrum of astrophysical
  neutrinos}},'' \href{http://dx.doi.org/10.1103/PhysRevD.95.083009}{Phys. Rev.
  {\normalfont \bfseries D95} (2017) no.~8, 083009},
\href{http://arxiv.org/abs/1611.07905}{{\normalfont \ttfamily
  arXiv:1611.07905}}.

\bibitem{Kistler:2016ask}
M.~D. Kistler and R.~Laha, ``{\em {Multi-PeV Signals from a New Astrophysical
  Neutrino Flux Beyond the Glashow Resonance}},''
  \href{http://dx.doi.org/10.1103/PhysRevLett.120.241105}{Phys. Rev. Lett.
  {\normalfont \bfseries 120} (2018) no.~24, 241105},
\href{http://arxiv.org/abs/1605.08781}{{\normalfont \ttfamily
  arXiv:1605.08781}}.

\bibitem{Biehl:2016psj}
D.~Biehl, A.~Fedynitch, A.~Palladino, T.~J. Weiler, and W.~Winter, ``{\em
  {Astrophysical Neutrino Production Diagnostics with the Glashow
  Resonance}},'' \href{http://dx.doi.org/10.1088/1475-7516/2017/01/033}{JCAP
  {\normalfont \bfseries 1701} (2017)  033},
\href{http://arxiv.org/abs/1611.07983}{{\normalfont \ttfamily
  arXiv:1611.07983}}.

\bibitem{Sahu:2016qet}
S.~Sahu and B.~Zhang, ``{\em {On the non-detection of Glashow resonance in
  IceCube}},'' \href{http://dx.doi.org/10.1016/j.jheap.2018.01.003}{JHEAp
  {\normalfont \bfseries 18} (2018)  1--4},
  \href{http://arxiv.org/abs/1612.09043}{{\normalfont \ttfamily
  arXiv:1612.09043}}.

\bibitem{Huang:2019hgs}
G.-y. Huang and Q.~Liu, ``{\em {Hunting the Glashow Resonance with PeV Neutrino
  Telescopes}},'' \href{http://dx.doi.org/10.1088/1475-7516/2020/03/005}{JCAP
  {\normalfont \bfseries 03} (2020)  005},
  \href{http://arxiv.org/abs/1912.02976}{{\normalfont \ttfamily
  arXiv:1912.02976}}.

\bibitem{Zhou:2020oym}
S.~Zhou, ``{\em {Cosmic Flavor Hexagon for Ultrahigh-energy Neutrinos and
  Antineutrinos at Neutrino Telescopes}},''
  \href{http://arxiv.org/abs/2006.06181}{{\normalfont \ttfamily
  arXiv:2006.06181}}.

\bibitem{IceCube:2015fuw}
{\normalfont \bfseries IceCube}, M.~G. Aartsen {\em et al.}, ``{\em {The
  IceCube Neutrino Observatory - Contributions to ICRC 2015 Part II:
  Atmospheric and Astrophysical Diffuse Neutrino Searches of All Flavors}},''
  in {\em {34th International Cosmic Ray Conference}}.
\newblock 10, 2015.
\newblock \href{http://arxiv.org/abs/1510.05223}{{\normalfont \ttfamily
  arXiv:1510.05223}}.

\bibitem{Gauld:2019pgt}
R.~Gauld, ``{\em {Precise predictions for multi-TeV and PeV energy neutrino
  scattering rates}},''
  \href{http://dx.doi.org/10.1103/PhysRevD.100.091301}{Phys. Rev. {\normalfont
  \bfseries D100} (2019) no.~9, 091301},
\href{http://arxiv.org/abs/1905.03792}{{\normalfont \ttfamily
  arXiv:1905.03792}}.

\bibitem{Garcia:2020jwr}
A.~Garcia, R.~Gauld, A.~Heijboer, and J.~Rojo, ``{\em {Complete predictions for
  high-energy neutrino propagation in matter}},''
  \href{http://dx.doi.org/10.1088/1475-7516/2020/09/025}{JCAP {\normalfont
  \bfseries 09} (2020)  025},
  \href{http://arxiv.org/abs/2004.04756}{{\normalfont \ttfamily
  arXiv:2004.04756}}.

\bibitem{Alikhanov:2009vcg}
I.~Alikhanov, ``{\em {On parton distributions in a photon gas}},''
  \href{http://dx.doi.org/10.1140/epjc/s10052-009-1192-y}{Eur. Phys. J. C
  {\normalfont \bfseries 65} (2010)  269--273},
  \href{http://arxiv.org/abs/0812.0937}{{\normalfont \ttfamily
  arXiv:0812.0937}}.

\bibitem{Glashow:2014}
A.~Loewy, S.~Nussinov, and S.~L. Glashow, ``{\em {The Effect of Doppler
  Broadening on the $6.3 \ PeV$ $W^-$ Resonance in $\bar{\nu}_e e^-$
  Collisions}},''
\href{http://arxiv.org/abs/1407.4415}{{\normalfont \ttfamily arXiv:1407.4415}}.

\bibitem{ALEPH:2005ab}
{\normalfont \bfseries ALEPH, DELPHI, L3, OPAL, SLD, LEP Electroweak Working
  Group, SLD Electroweak Group, SLD Heavy Flavour Group}, S.~Schael {\em et
  al.}, ``{\em {Precision electroweak measurements on the $Z$ resonance}},''
  \href{http://dx.doi.org/10.1016/j.physrep.2005.12.006}{Phys. Rept.
  {\normalfont \bfseries 427} (2006)  257--454},
  \href{http://arxiv.org/abs/hep-ex/0509008}{{\normalfont \ttfamily
  arXiv:hep-ex/0509008}}.

\bibitem{Cacciari:1992}
M.~Cacciari, A.~Deandrea, G.~Montagna, and O.~Nicrosini, ``{\em {QED} Structure
  Functions: A Systematic Approach},''
  \href{http://dx.doi.org/10.1209/0295-5075/17/2/007}{Europhysics Letters
  ({EPL}) {\normalfont \bfseries 17} (1992) no.~2, 123--128}.

\bibitem{IceCube:2014stg}
{\normalfont \bfseries IceCube}, M.~G. Aartsen {\em et al.}, ``{\em
  {Observation of High-Energy Astrophysical Neutrinos in Three Years of IceCube
  Data}},'' \href{http://dx.doi.org/10.1103/PhysRevLett.113.101101}{Phys. Rev.
  Lett. {\normalfont \bfseries 113} (2014)  101101},
  \href{http://arxiv.org/abs/1405.5303}{{\normalfont \ttfamily
  arXiv:1405.5303}}.

\bibitem{Hillas:1984}
A.~M. Hillas, ``{\em The Origin of Ultra-High-Energy Cosmic Rays},''
  \href{http://dx.doi.org/10.1146/annurev.aa.22.090184.002233}{Annual Review of
  Astronomy and Astrophysics {\normalfont \bfseries 22} (1984) no.~1,
  425--444}.

\bibitem{Cowan:1998ji}
G.~Cowan, {\em {Statistical data analysis}}.
\newblock Clarendon Press, 1998.

\bibitem{Baerwald_2011}
P.~Baerwald, S.~Hümmer, and W.~Winter, ``{\em Magnetic field and flavor
  effects on the gamma-ray burst neutrino flux},'' Phys. Rev. D {\normalfont
  \bfseries 83} (2011) no.~6, ,
  \href{http://arxiv.org/abs/1009.4010}{{\normalfont \ttfamily
  arXiv:1009.4010}}.

\bibitem{Huemmer_2010}
S.~Hümmer, M.~Rüger, F.~Spanier, and W.~Winter, ``{\em Simplified models for
  photohadronic interactions in cosmic accelerators},'' Astrophys. J.
  {\normalfont \bfseries 721} (2010) no.~1, 630--652,
  \href{http://arxiv.org/abs/1002.1310}{{\normalfont \ttfamily
  arXiv:1002.1310}}.

\bibitem{IceCube:2014gqr}
{\normalfont \bfseries IceCube}, M.~G. Aartsen {\em et al.}, ``{\em
  {IceCube-Gen2: A Vision for the Future of Neutrino Astronomy in
  Antarctica}},'' \href{http://arxiv.org/abs/1412.5106}{{\normalfont \ttfamily
  arXiv:1412.5106}}.

\bibitem{IceCube-Gen2:2020qha}
{\normalfont \bfseries IceCube-Gen2}, M.~G. Aartsen {\em et al.}, ``{\em
  {IceCube-Gen2: the window to the extreme Universe}},''
  \href{http://dx.doi.org/10.1088/1361-6471/abbd48}{J. Phys. G {\normalfont
  \bfseries 48} (2021) no.~6, 060501},
  \href{http://arxiv.org/abs/2008.04323}{{\normalfont \ttfamily
  arXiv:2008.04323}}.

\bibitem{Baikal-GVD:2018isr}
{\normalfont \bfseries Baikal-GVD}, A.~D. Avrorin {\em et al.}, ``{\em
  {Baikal-GVD: status and prospects}},''
  \href{http://dx.doi.org/10.1051/epjconf/201819101006}{EPJ Web Conf.
  {\normalfont \bfseries 191} (2018)  01006},
  \href{http://arxiv.org/abs/1808.10353}{{\normalfont \ttfamily
  arXiv:1808.10353}}.

\bibitem{KM3Net:2016zxf}
{\normalfont \bfseries KM3Net}, S.~Adrian-Martinez {\em et al.}, ``{\em {Letter
  of intent for KM3NeT 2.0}},''
  \href{http://dx.doi.org/10.1088/0954-3899/43/8/084001}{J. Phys. G
  {\normalfont \bfseries 43} (2016) no.~8, 084001},
  \href{http://arxiv.org/abs/1601.07459}{{\normalfont \ttfamily
  arXiv:1601.07459}}.

\bibitem{P-ONE:2020ljt}
{\normalfont \bfseries P-ONE}, M.~Agostini {\em et al.}, ``{\em {The Pacific
  Ocean Neutrino Experiment}},''
  \href{http://dx.doi.org/10.1038/s41550-020-1182-4}{Nature Astron.
  {\normalfont \bfseries 4} (2020) no.~10, 913--915},
  \href{http://arxiv.org/abs/2005.09493}{{\normalfont \ttfamily
  arXiv:2005.09493}}.

\bibitem{Romero-Wolf:2020pzh}
A.~Romero-Wolf {\em et al.}, ``{\em {An Andean Deep-Valley Detector for
  High-Energy Tau Neutrinos}},'' in {\em {Latin American Strategy Forum for
  Research Infrastructure}}.
\newblock 2, 2020.
\newblock \href{http://arxiv.org/abs/2002.06475}{{\normalfont \ttfamily
  arXiv:2002.06475}}.

\bibitem{Ye:2022vbk}
Z.~P. Ye {\em et al.}, ``{\em {Proposal for a neutrino telescope in South China
  Sea}},'' \href{http://arxiv.org/abs/2207.04519}{{\normalfont \ttfamily
  arXiv:2207.04519}}.

\bibitem{Huang:2021mki}
G.-y. Huang, S.~Jana, M.~Lindner, and W.~Rodejohann, ``{\em {Probing new
  physics at future tau neutrino telescopes}},''
  \href{http://dx.doi.org/10.1088/1475-7516/2022/02/038}{JCAP {\normalfont
  \bfseries 02} (2022) no.~02, 038},
  \href{http://arxiv.org/abs/2112.09476}{{\normalfont \ttfamily
  arXiv:2112.09476}}.

\bibitem{Coleman:2022abf}
A.~Coleman {\em et al.}, ``{\em {Ultra high energy cosmic rays The intersection
  of the Cosmic and Energy Frontiers}},''
  \href{http://dx.doi.org/10.1016/j.astropartphys.2022.102794}{Astropart. Phys.
  {\normalfont \bfseries 147} (2023)  102794},
  \href{http://arxiv.org/abs/2205.05845}{{\normalfont \ttfamily
  arXiv:2205.05845}}.

\bibitem{Ackermann:2022rqc}
M.~Ackermann {\em et al.}, ``{\em {High-energy and ultra-high-energy neutrinos:
  A Snowmass white paper}},''
  \href{http://dx.doi.org/10.1016/j.jheap.2022.08.001}{JHEAp {\normalfont
  \bfseries 36} (2022)  55--110},
  \href{http://arxiv.org/abs/2203.08096}{{\normalfont \ttfamily
  arXiv:2203.08096}}.

\bibitem{Valera:2022wmu}
V.~B. Valera, M.~Bustamante, and C.~Glaser, ``{\em {Near-future discovery of
  the diffuse flux of ultra-high-energy cosmic neutrinos}},''
  \href{http://arxiv.org/abs/2210.03756}{{\normalfont \ttfamily
  arXiv:2210.03756}}.

\bibitem{LHAASO:2021cbz}
{\normalfont \bfseries LHAASO*\textdagger{}, LHAASO}, Z.~Cao {\em et al.},
  ``{\em {Peta\textendash{}electron volt gamma-ray emission from the Crab
  Nebula}},'' \href{http://dx.doi.org/10.1126/science.abg5137}{Science
  {\normalfont \bfseries 373} (2021) no.~6553, 425--430},
  \href{http://arxiv.org/abs/2111.06545}{{\normalfont \ttfamily
  arXiv:2111.06545}}.

\bibitem{LHAASO_nature}
Z.~Cao {\em et al.}, ``{\em {Ultrahigh-energy photons up to 1.4
  petaelectronvolts from 12 $\gamma$-ray Galactic sources}},'' Nature
  {\normalfont \bfseries 594} (2021) no.~7861, 33--36.

\bibitem{Sudoh:2022sdk}
T.~Sudoh and J.~F. Beacom, ``{\em {Where are Milky~Way\textquoteright{}s
  hadronic PeVatrons?}},''
  \href{http://dx.doi.org/10.1103/PhysRevD.107.043002}{Phys. Rev. D
  {\normalfont \bfseries 107} (2023) no.~4, 043002},
  \href{http://arxiv.org/abs/2209.03970}{{\normalfont \ttfamily
  arXiv:2209.03970}}.

\bibitem{Bustamante:2020niz}
M.~Bustamante, ``{\em {New limits on neutrino decay from the Glashow resonance
  of high-energy cosmic neutrinos}},''
  \href{http://arxiv.org/abs/2004.06844}{{\normalfont \ttfamily
  arXiv:2004.06844}}.

\bibitem{Jezo:2014kla}
T.~Ježo, M.~Klasen, F.~Lyonnet, F.~Montanet, I.~Schienbein, and M.~Tartare,
  ``{\em {Can new heavy gauge bosons be observed in ultra-high energy cosmic
  neutrino events?}},''
  \href{http://dx.doi.org/10.1103/PhysRevD.89.077702}{Phys. Rev. {\normalfont
  \bfseries D89} (2014) no.~7, 077702},
\href{http://arxiv.org/abs/1401.6012}{{\normalfont \ttfamily arXiv:1401.6012}}.

\bibitem{Babu:2019vff}
K.~S. Babu, P.~S. Dev, S.~Jana, and Y.~Sui, ``{\em {Zee-Burst: A New Probe of
  Neutrino Non-Standard Interactions at IceCube}},''
\href{http://arxiv.org/abs/1908.02779}{{\normalfont \ttfamily
  arXiv:1908.02779}}.

\bibitem{Dey:2020fbx}
U.~K. Dey, N.~Nath, and S.~Sadhukhan, ``{\em {Charged Higgs effects in IceCube:
  PeV events and NSIs}},''
  \href{http://dx.doi.org/10.1007/JHEP09(2021)113}{JHEP {\normalfont \bfseries
  09} (2021)  113}, \href{http://arxiv.org/abs/2010.05797}{{\normalfont
  \ttfamily arXiv:2010.05797}}.

\bibitem{Babu:2022fje}
K.~S. Babu, P.~S.~B. Dev, and S.~Jana, ``{\em {Probing neutrino mass models
  through resonances at neutrino telescopes}},''
  \href{http://dx.doi.org/10.1142/S0217751X22300034}{Int. J. Mod. Phys. A
  {\normalfont \bfseries 37} (2022) no.~11n12, 2230003},
  \href{http://arxiv.org/abs/2202.06975}{{\normalfont \ttfamily
  arXiv:2202.06975}}.

\bibitem{Xu:2022svm}
D.-H. Xu and S.-J. Rong, ``{\em {Connect the Lorentz Violation to the Glashow
  Resonance Event}},'' \href{http://arxiv.org/abs/2211.05478}{{\normalfont
  \ttfamily arXiv:2211.05478}}.

\bibitem{Arguelles:2022xxa}
C.~A. Arg\"uelles {\em et al.}, ``{\em {Snowmass White Paper: Beyond the
  Standard Model effects on Neutrino Flavor}},'' in {\em {2022 Snowmass Summer
  Study}}.
\newblock 3, 2022.
\newblock \href{http://arxiv.org/abs/2203.10811}{{\normalfont \ttfamily
  arXiv:2203.10811}}.

\bibitem{Huang:2022pce}
G.-y. Huang, S.~Jana, M.~Lindner, and W.~Rodejohann, ``{\em {Probing Heavy
  Sterile Neutrinos at Ultrahigh Energy Neutrino Telescopes via the Dipole
  Portal}},'' \href{http://arxiv.org/abs/2204.10347}{{\normalfont \ttfamily
  arXiv:2204.10347}}.

\bibitem{Huang:2022ebg}
G.-y. Huang, ``{\em {Double and multiple bangs at tau neutrino telescopes}},''
  \href{http://dx.doi.org/10.1140/epjc/s10052-022-11052-y}{Eur. Phys. J. C
  {\normalfont \bfseries 82} (2022) no.~12, 1089},
  \href{http://arxiv.org/abs/2207.02222}{{\normalfont \ttfamily
  arXiv:2207.02222}}.

\bibitem{Heighton:2023qpg}
R.~Heighton, L.~Heurtier, and M.~Spannowsky, ``{\em {Hunting for Neutral
  Leptons with Ultra-High-Energy Cosmic Rays}},''
  \href{http://arxiv.org/abs/2303.11352}{{\normalfont \ttfamily
  arXiv:2303.11352}}.

\bibitem{Clementi:1963}
E.~Clementi and D.~L. Raimondi, ``{\em {Atomic Screening Constants from SCF
  Functions}},'' \href{http://dx.doi.org/10.1063/1.1733573}{jcp {\normalfont
  \bfseries 38} (1963) no.~11, 2686--2689}.

\end{thebibliography}\endgroup

\end{document}